\documentclass[universe,article,accept,pdftex,moreauthors]{Definitions/mdpi}   

\usepackage{graphicx} 
\usepackage{color}    
\usepackage{float}
\usepackage[justification=centering]{caption}
\usepackage{placeins}
\usepackage{multicol}
\usepackage{amsmath}
\usepackage{verbatim}
\usepackage{array}
\usepackage{chemist}
\usepackage{soul}
\usepackage{booktabs}
\usepackage{threeparttable}
\usepackage{amssymb}% http://ctan.org/pkg/amssymb
\usepackage{pifont}% http://ctan.org/pkg/pifont
\usepackage{array,multirow}
\usepackage{tikz}
\usetikzlibrary{shapes,backgrounds,calc}

\makeatletter
\tikzset{circle split part fill/.style  args={#1,#2}{%
 alias=tmp@name, %
  postaction={%
    insert path={
     \pgfextra{% 
     \pgfpointdiff{\pgfpointanchor{\pgf@node@name}{center}}%
                  {\pgfpointanchor{\pgf@node@name}{east}}%            
     \pgfmathsetmacro\insiderad{\pgf@x}
      \fill[#1] (\pgf@node@name.base) ([xshift=-\pgflinewidth]\pgf@node@name.east) arc
                          (0:180:\insiderad-\pgflinewidth)--cycle;
      \fill[#2] (\pgf@node@name.base) ([xshift=\pgflinewidth]\pgf@node@name.west)  arc
                           (180:360:\insiderad-\pgflinewidth)--cycle;            %  \end{scope}   
         }}}}}  
 \makeatother  

\def\be{\begin{equation}} 
\def\ee{\end{equation}}

\def\msun{{\Msun}}

\def\gsim{\lower.5ex\hbox{\gtsima}} 
\def\lsim{\lower.5ex\hbox{\ltsima}} \def\gtsima{$\; \buildrel > \over 
\sim \;$} \def\ltsima{$\; \buildrel < \over \sim \;$} \def\prosima{$\; 
\buildrel \propto \over \sim \;$} \def\gsim{\lower.5ex\hbox{\gtsima}} 
\def\lsim{\lower.5ex\hbox{\ltsima}} 
\def\simgt{\lower.5ex\hbox{\gtsima}} 
\def\simlt{\lower.5ex\hbox{\ltsima}} 
\def\simpr{\lower.5ex\hbox{\prosima}}   
  
 \def\gtsima{$\; \buildrel > \over \sim \;$} 
\def\ltsima{$\; \buildrel < \over \sim \;$} 
\def\gsim{\lower.5ex\hbox{\gtsima}} 
\def\lsim{\lower.5ex\hbox{\ltsima}} 
\def\simgt{\lower.5ex\hbox{\gtsima}} 
\def\simlt{\lower.5ex\hbox{\ltsima}} 
\def\simpr{\lower.5ex\hbox{\prosima}}

\def\msun{\,{\rm \Msun}}

\def\E3{{\cal E}_{\rm g}^{III}}

\def\Msun{\rm M_\odot}

\def\Msun{\rm M_\odot}

\def\M*{M_*}
\def\Z*{Z_*}
\def\L*{L_*}

\tikzstyle{igm} = [rectangle, rounded corners, minimum width=2cm, minimum height=1cm,text centered, draw=black, fill=gray!30]
\tikzstyle{hgas} = [rectangle, rounded corners, minimum width=2cm, minimum height=1cm,text centered, draw=black, fill=red!30]
\tikzstyle{cgas} = [rectangle, rounded corners, minimum width=2cm, minimum height=1cm,text centered, draw=black, fill=blue!30]
\tikzstyle{stars} = [rectangle, rounded corners, minimum width=2cm, minimum height=1cm,text centered, draw=black, fill=yellow!30]
\tikzstyle{bh} = [rectangle, rounded corners, minimum width=2cm, minimum height=1cm,text centered, draw=black, fill=black!80]
\tikzstyle{res} = [rectangle, rounded corners, minimum width=2cm, minimum height=1cm,text centered, draw=black, fill=green!40]
\tikzstyle{box} = [rectangle, rounded corners, minimum width=2.8cm, minimum height=2.8cm,text centered, draw=black]
\tikzstyle{box2} = [rectangle, rounded corners, minimum width=2.8cm, minimum height=1.5cm,text centered, draw=black, fill=gray!30]
\tikzstyle{box1} = [rectangle, rounded corners, minimum width=1.5cm, minimum height=1cm,text centered, draw=none]
\tikzstyle{arrow} = [thick,->,>=stealth]
\tikzstyle{prog1} = [circle, minimum size=1.5cm,text centered, draw=black, fill=black!25]
\tikzstyle{prog2} = [circle, minimum size=2.2cm,text centered, draw=black, fill=black!25]
\tikzstyle{prog3} = [circle, minimum size=1.9cm,text centered, draw=black, fill=black!25]
\tikzstyle{prog4} = [circle, minimum size=2.5cm,text centered, draw=black, fill=black!25]
\tikzstyle{split1} = [circle split, minimum size=1.3cm, line width=0pt, circle split part fill={blue!30,red!30}]
\tikzstyle{split2} = [circle split, minimum size=2cm, circle split part fill={blue!30,red!30}]
\tikzstyle{split3} = [circle split, minimum size=1.7cm, circle split part fill={blue!30,red!30}]
\tikzstyle{split4} = [circle split, minimum size=2.3cm, circle split part fill={blue!30,red!30}]
\tikzstyle{star1} = [star, star points=8, minimum size=1cm, star point height=2mm, fill=yellow]
\tikzstyle{star2} = [star, star points=8, minimum size=1.5cm, star point height=2mm, fill=yellow]
\tikzstyle{star3} = [star, star points=8, minimum size=1.2cm, star point height=2mm, fill=yellow]
\tikzstyle{star4} = [star, star points=8, minimum size=1.8cm, star point height=2mm, fill=yellow]
\tikzstyle{bh1} = [circle, minimum size=0.3cm,text centered, draw=black, fill=black]
\tikzstyle{bh2} = [circle, minimum size=0.6cm,text centered, draw=black, fill=black]
\tikzstyle{bh3} = [circle, minimum size=0.4cm,text centered, draw=black, fill=black]
\tikzstyle{bh4} = [circle, minimum size=0.8cm,text centered, draw=black, fill=black]

\firstpage{1} 
\pubvolume{1}
\issuenum{1}
\articlenumber{0}
\pubyear{2024}
\copyrightyear{2024}
\externaleditor{Academic Editor: Firstname Lastname}
\datereceived{21 January 2025} 
\daterevised{18 February 2025} % Comment out if no revised date
\dateaccepted{21 February 2025} 
\datepublished{ } 
\hreflink{https://doi.org/} % If needed use \linebreak
\Title{Ultra-High-Energy Cosmic Rays from Active Galactic Nuclei Jets%EE - Attention AE - Title change
	: The Role of Supermassive Black Hole Growth and Accretion States}

\TitleCitation{Ultra-High-Energy Cosmic Rays from Active Galactic Nuclei Jets: The Role of Supermassive Black Hole Growth and Accretion States}

\Author{Olmo Piana %MDPI: 1. Please carefully check the accuracy of names and affiliations. 2. For author from Taiwan, the preferred format is “Xiao-Ming Wang”; the name format should be as uniform as possible. Please check and confirm.
 $^{1,2,}$*%MDPI: we added asterisk to indicated correspondence author, please check and confirm.
\orcidA{}, Hung-Yi Pu $^{1,2,3,}$*\orcidB{}}
\AuthorNames{Olmo Piana, Hung-Yi Pu}
\isAPAStyle{%
       \AuthorCitation{Lastname, F., Lastname, F., \& Lastname, F.}
         }{%
        \isChicagoStyle{%
        \AuthorCitation{Lastname, Firstname, Firstname Lastname, and Firstname Lastname.}
        }{
        \AuthorCitation{Piana, O.; Pu, H.-Y.}
        }
}

\address{%
$^{1}$ \quad Department of Physics, National Taiwan Normal University, No. 88,  Section 4, Tingzhou Road, \linebreak Taipei 116, Taiwan %MDPI: We revised country name format, please confirm.
\\
$^{2}$ \quad Centre of Astronomy and Gravitation, National Taiwan Normal University, No. 88,  Section 4, Tingzhou Road, Taipei 116, Taiwan\\
$^{3}$ \quad Institute of Astronomy and Astrophysics, Academia Sinica, 11F of Astronomy-Mathematics Building, AS/NTU No. 1, Section %MDPI: Please check if this should be `Section 4.' please keep consistent format throughout the paper.
 4, Roosevelt Road%MDPI: Please check if this should be `Road' please keep consistent format throughout the paper.
, Taipei 10617, Taiwan}

\corres{Correspondence: piana@ntnu.edu.tw (O.P.); hypu@gapps.ntnu.edu.tw (H.-Y.P.)}
\abstract{Jets emanating from active galactic nuclei (AGN) represent some of the most formidable particle accelerators in the universe, thereby emerging as viable candidates responsible for the detection of ultra-high-energy cosmic rays (UHECRs). If AGN jets indeed serve as origins of UHECRs, then the diffuse flux of these cosmic rays would be dependent on the power and duty cycle of these jets, which are inherently connected to the nature of black hole accretion flows. In this article, we present our cosmological semi-analytic framework, \textit{JET}%MDPI: Please confirm if the italics is unnecessary and can be removed. Please check throughout the whole paper
 (Jets from Early Times), designed to trace the evolution of jetted AGN populations. This framework serves as a valuable tool for predictive analyses of cosmic ray energy density and, potentially, neutrino energy density. By using \textit{JET}, we model the formation and evolution of galaxies and supermassive black holes (SMBHs) from $z=20$ to $z=1$, incorporating jet formation and feedback mechanisms and distinguishing between various accretion states determined by the SMBH Eddington ratios. The implications of different SMBH growth models on predicting cosmic ray flux are investigated. We provide illustrative examples demonstrating how the associated diffuse UHECR fluxes \textcolor{black}{at the source} may vary in relation to the jet production efficiencies and the selected SMBH growth model, linking cosmological models of SMBH growth with astroparticle backgrounds.}


\begin{document}

\section{Introduction}
The origin of ultra-high-energy cosmic rays (UHECRs) with energies above $10^{17}$ eV, presumed to come from extragalactic sources, continues to be a subject of investigation. Several works have shown that UHERCs can play an important role in galaxy \mbox{evolution \citep{naab2017, owen2019, buck2020, semenov2021, owen2023, ruszkowski2023}}, affecting the morphology and dynamics of molecular clouds in star-forming galaxies by depositing part of their energy into the interstellar medium (ISM) \citep{indriolo2009, casanova2010, owen2021, gabici2022} and potentially driving gas outflows \citep{ipavich1975, yu2021, thompson2024}. On~an intergalactic scale, after~escaping their galaxy of origin, they can interact with the warm-hot intergalactic medium present in filaments \citep{wu2024} and~be transported around the cosmic web to other galaxies. For~all these reasons, we believe that incorporating CR physics into cosmological models of galaxy evolution is of significant importance. The~origin of cosmic rays, however, remains elusive, and there are considerable challenges in constraining source population models using ground-based observations of the CR flux \citep{kotera2008, takami2011, kachelriess2019, alvesbatista2019, owen2023a}. The~process of diffusive particle acceleration occurring behind the shock fronts of supernovae (SNe) \citep{blasi2010} as well as within the coronae of active galactic nuclei (AGNs) \citep{inoue2019} may account for the acceleration of cosmic rays up to energies of $\sim$3 $\times 10^{15}$ eV, potentially reaching up to $10^{17}$ eV for the heaviest nuclei when strong magnetic fields are present. This acceleration mechanism is insufficient, however, to~account for the UHE tail of the CR spectrum, which extends up to $\sim$10$^{20}$ eV. Relativistic jets in AGNs are often considered a viable site for UHECR acceleration due to their immense power (e.g., %MDPI: We revised reference citation to this format, please confirm the revision.
 \citep{caprioli2015}, and references therein), while alternative energetic phenomena, including gamma ray bursts and tidal disruption events, have also been considered as potential contributors \citep{bykov2019, globus2023, zirakashvili2024}. Initial observations conducted by the Pierre Auger Collaboration appeared to support the AGN hypothesis, as~they detected spatial anisotropies in the UHECR field that seem to align with the positions of AGN in the sky \citep{pac2007}; however, the~statistical significance of these findings remains a topic of debate \citep{aab2015}. \textcolor{black}{Observations carried out with Cherenkov telescopes such as HESS, MAGIC, and VERITAS, able to observe diffuse $\gamma$-ray fluxes, also  suggest that active SMBHs can provide an origin for PeV and ultra-high-energy cosmic rays \citep{abramowski2016, acciari2020, toomey2020}, confirming at the same time that SNe remnants alone are not sufficient to produce them \citep{archambault2017, abeysekara2020}.}

Black hole jet activity, found in systems of both stellar-mass black holes in black hole X-ray binaries (BHXBs) and SMBHs in AGN, is associated with the accretion rate and therefore the type of accretion (see, e.g., \cite{mirabel2009}). For~example, according to their hardness in X-ray spectra and total intensity, BHXBs exhibit jet activity at a low/hard state: when the X-ray spectrum is harder and the total intensity (and also the corresponding accretion rate) is lower. In~contrast, the~jet quenches when the system transits to a high/soft state: the X-ray spectrum is now softer and the total intensity is higher (e.g., \citep{fender2004,belloni2010}). Similar spectral \mbox{states \citep{sobolewska2011,moravec2022}} and disk-jet couplings \citep{trump2011} have also been suggested for AGN systems. A~typical threshold for state transition and jet suppression is when the Eddington ratio grows above $\lambda_\mathrm{ADAF,jet}\sim 0.001--0.01$, \textcolor{black}{where $\lambda\equiv\dot{M}/(\dot{M}_{\rm Edd})$ and the Eddington accretion rate, $\dot{M}_{\rm Edd}$, is related to the Eddington luminosity, $L_{\rm Edd}$, with $\eta\dot{M}_{\rm Edd}c^{2}=L_{\rm Edd}$, $\eta=0.1$}. Theoretically, below~this ratio the central region is occupied by a radiatively inefficient geometrically thick and optically thin accretion flow (ADAF, advection-dominated accretion flow) \cite{ichimaru1977,narayan1994,narayan1995}. When the Eddington ratio increases above $\lambda_\mathrm{ADAF,jet}$, the~accretion disc becomes optically thick, it can efficiently radiate away the accumulated energy, thus remaining cool and geometrically thin. The~system enters in the thermal state and the jets are suppressed. At~even higher accretion rates, for~$\lambda_\mathrm{slim,jet} \gtrsim 1$, the~accretion becomes geometrically thick again because the photons are trapped inside the accretion flow, resulting in a slim disc \mbox{(see, e.g., \cite{sadowski2009}). }The~magnetic extraction of rotational energy from black holes is thought to be responsible for the launching mechanisms of both ADAF jets and slim-disc jets. However, in~the latter cases, the~process can be more complex due to the significant influence exerted by radiation pressure (see, e.g., \citep{sadowski2016}). Despite the current incomplete understanding of the distinctions between these two types of jet, it is expected that the associated production of UHECRs from these jets may~vary.

Ref. %MDPI: The Ref. number is not allowed to be used as the subject in sentences of the main text, we added `Ref.' before ref. citation, please confirm the revision. The following highlights are the same.
 \citep{piana2024} (from now on referred to as PPW24) have modeled the historical activity of supermassive black hole jets from $z=20$ to $z=4$, encompassing a comprehensive range of halo masses while considering both types of jet: those operating at low accretion rates (ADAF jets) and those at high accretion rates (slim-disc jets). Two distinct accretion scenarios are evaluated: (1) super-Eddington model (sEDD; fiducial model in PPW24) and~(2) Eddington-limited model (EDDlim; reference model in PPW24). Both scenarios effectively replicate the primary statistical properties of AGN and galaxies, as~reported in $z \lsim 7$; their predictions, conversely, differ substantially at higher redshifts and for black holes in the intermediate mass range $10^4 \lsim M_\mathrm{bh} \lsim 10^6 \mathrm{M_\odot}$, enabling us to potentially distinguish between the two black hole growth~models.

This study investigates the predicted energy fluxes and spectral properties of UHECRs \textcolor{black}{at the source} by incorporating our cosmological model of galaxy and SMBH formation and evolution, as~proposed in PPW24, with~models for the production of cosmic rays in AGN jets.
We examine two distinct models for the production of UHECRs: in the first model, UHECRs are exclusively generated by ADAF jets, whereas in the second model both ADAF and slim-disc jets produce UHECRs with the same efficiency. Furthermore, in~contrast to PPW24, the~cosmic evolution from $z=20$ to  $z=1$ is modeled, along with a consideration of metallicity evolution. \textcolor{black}{It is crucial to consider that the predicted UHECR spectrum is fundamentally dependent on the number density and power of the jetted AGN at specific redshifts, and~consequently it is influenced by the associated uncertainties due to the limited constraints on jet properties at high redshifts. However, SMBH growth and feedback models during early epochs are expected to be constrained more effectively through the surveys conducted by current and forthcoming missions such as JWST \citep{gardner2006},  Laser Space Interferometer Antenna (LISA), and~Advanced Telescope for High-energy Astrophysics (ATHENA).}

The article is organized as follows: In Section~\ref{model}, we recap the characteristics of the model, with~a special focus on the features that are new with respect to the implementation in PPW24. In~Section~\ref{results_1}, we show our predictions for the main AGN and galaxy observables to make sure that our model is properly tuned. In~Section~\ref{results_2}, we introduce our formula for UHECR production in AGN jets and show the results with respect to different scenarios. We conclude with a summary and discussion in Section~\ref{conclusions}.

%%%%%%%%%%%%%%%%%%%%%%%%%%%%%%%%%%%%%%%%%%
 \section{The \textit{JET} Model for Galaxy and Black Hole~Evolution}
\label{model}

Our cosmological semi-analytic framework, \textit{JET} (Jets from Early Times), finds its roots in the DELPHI cosmological framework \citep{dayal2014, dayal2019, piana2021, piana2022}, yet has been adapted to incorporate a model of jet formation (as already applied in PPW24) and to reach lower redshifts. 
In this section, we outline the primary characteristics of our model and present a comprehensive examination of the recent modifications implemented. For~an overview of the parameters adopted in the model, see Table~\ref{table_params} in Appendix~\ref{appendixA}.

\subsection{Dark Matter Merger~Tree}
\begin{figure}[H]
%\centering
\includegraphics[width=0.8\textwidth]{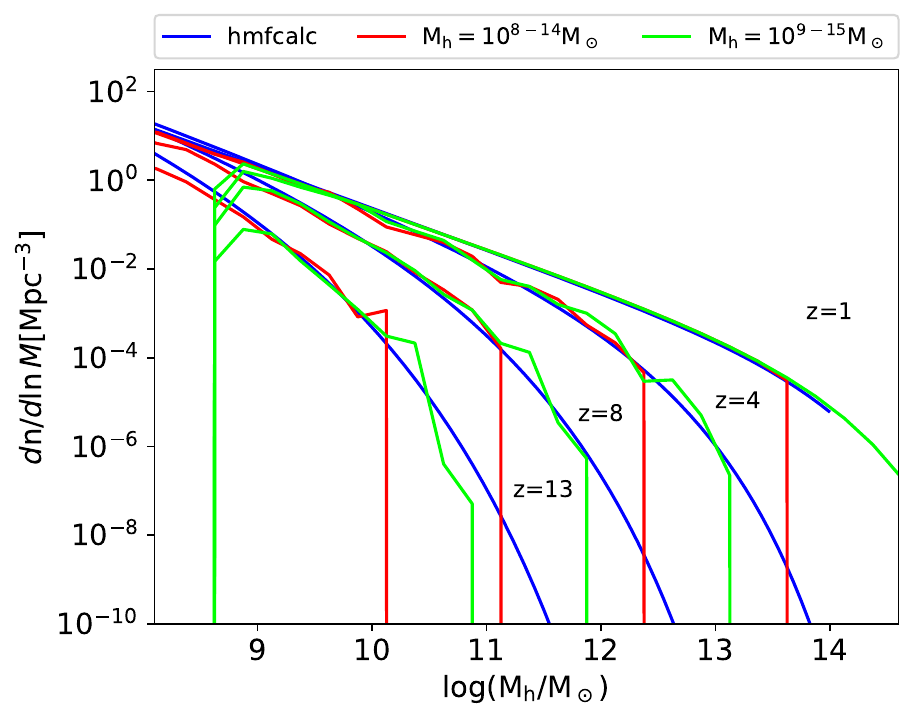}
\caption{Merger trees at different redshift. The~blue lines show the Sheth--Tormen halo mass function from the online published tool HMFcalc \protect\citep{murray2013}. Red lines correspond to our merger trees from halos within the $10^8-10^{14} \mathrm{M_\odot}$ mass range and a $10^8 \mathrm{M_\odot}$ mass resolution, while green lines cover the $10^9-10^{15} \mathrm{M_\odot}$ range with a $10^9 \mathrm{M_\odot}$ mass resolution (used in this work).}\label{mergert}
\end{figure}

\textcolor{black}{A dark matter merger tree of 120 halos with final masses $M_\mathrm{h} = 10^9-10^{15} \mathrm{M_\odot}$ is used for our semi-analytic model of galaxy evolutions and black hole growth. 
From $z=20$ to $z=1$, the~merger trees are computed in time steps of 20 Myr and with a mass resolution of $10^9 \mathrm{M_\odot}$.
The number density of the final halos and their progenitors are assigned to be consistent with the Sheth--Tormen halo mass function (HMF \citep{sheth-tormen1999}).} 
Figure~\ref{mergert} shows that adopting a higher mass resolution does not pose a problem for the dark matter halo statistics. Merger trees, respectively with $10^8 \mathrm{M_\odot}$ (red lines) and $10^9 \mathrm{M_\odot}$ (green lines) mass resolutions, reproduce the theoretical Sheth--Tormen HMF (blue lines) equally well. 
The merger trees keep track of both halo mergers and dark matter accreted smoothly below the resolution mass threshold. \textcolor{black}{New halos form with a dark matter mass just above the mass resolution threshold, and~with a gas mass proportional to $\mathrm{\Omega_b/\Omega_m}$. At~each time step, beside receiving the contribution from mergers with other halos, each halo accretes dark matter and, proportionally, gas from the intergalactic space.} The initial leaves of the merger trees at $z>13$ are seeded with a $150\msun$ stellar black hole (e.g., see \citep {latif2016, inayoshi2020, volonteri2021} for a review on the formation channels of black hole seeds).
It is worth pointing out that the final results are not significantly affected by implementing alternative seeding scenarios \citep{dayal2019}, since SMBH growth in such a model turns out to be self-regulated: higher-mass initial seeds will grow slower than lower-mass seeds, due to more~feedback. 
\unskip   

%#################################################################

\subsection{Star Formation and the Gas~Cycle}
\label{model_overview}
\textcolor{black}{As described in PPW24, during~each discrete temporal iteration, $z$, the~variations in mass for the dark matter halo, hot and cold gas, black hole, stellar components, and~the halo's gaseous reservoir are calculated. Furthermore, it is assumed that an unresolved amount of dark matter and gas originating from intergalactic regions would contribute to each halo.
The gas dynamics also includes the gas accreted from the IGM, and~that ejected from the galaxy via feedback processes.} 
Gas from the IGM (and the gas reservoir) can be accreted either cold, if~the halo mass, $M_\mathrm{h}$, is lower than a critical value, $M_\mathrm{h}^\mathrm{crit}$, or~hot and cold if $M_\mathrm{h} > M_\mathrm{h}^\mathrm{crit}$. Consistently with \textcolor{black}{what was} found in~\cite{dekel2006}, however, at~$z<2$ halos more massive than $M_\mathrm{h}^\mathrm{crit}$ can only accrete hot gas. The~critical halo mass, which also defines the SMBH accretion model, as~we will see in the corresponding subsection, is treated as a \mbox{free parameter:}
\begin{equation}
    M_\mathrm{h}^\mathrm{crit}(z) = 10^{11.25} {\Delta_z}^{-3/8} \mathrm{M_\odot} \ .
    \label{mhcrit}
\end{equation} 
\begin{comment}
according to
\begin{equation}
M^\mathrm{acc}_\mathrm{dm}(z) = \left[M_\mathrm{h}(z) - \sum_j M_\mathrm{h}^j(z+\Delta z)\right],
\label{mdotdm}
\end{equation}
where the sum runs over all the $j$ progenitors at $z + \Delta z$. Together with the unresolved dark matter, the~galaxy accretes from the IGM a gas mass proportional to the cosmic baryonic fraction. The~progenitors of a halo will bring in also their content of gas, stars and black hole mass. We can then write
\begin{equation}
\begin{split}
&M^\mathrm{cold}(z) = \sum_j M_j^\mathrm{cold}(z+\Delta z) + \dot{M}^\mathrm{cold}(z) \tau_s\  ,\\
&M^\mathrm{hot}(z) = \sum_j M_j^\mathrm{hot}(z+\Delta z) + \dot{M}^\mathrm{hot}(z) \mathrm{\tau_s}\ , \\
&M_*(z) = \sum_j M^j_*(z+\Delta z) + \dot{M}_*(z) \tau_s\ ,\\
&M_\mathrm{bh}(z) = \sum_j M^j_\mathrm{bh}(z+\Delta z) + \dot{M}_\mathrm{bh}(z) \mathrm{\tau_s}\ ,
\end{split}
\label{step_ev}
\end{equation}
where the index $j$ runs over all the progenitors of the halo at the previous time step, and~$\tau_s$ is the time step employed in our model, which in our fiducial case is 20 Myr. In~the later subsections we will omit the dependence on $z$ from the equations, for~simplicity, and~we will assume the different terms all refers to the same time step $z$, unless~otherwise specified.
\end{comment}
At %MDPI: Please confirm if no-indent paragraph shoud be retained. The following highlights are the same..
each time step, part of the cold gas is turned into stars, and~star formation feedback drives gas outflows into the gas reservoir.
For more details on how the different gas phases and the stellar mass are tracked, we refer the reader to~PPW24. 

%*******************************************************************
\subsection{Metallicity~Evolution}
Stellar activity in galaxies produces metals that can then be ejected into the ISM and the gas reservoir. Stellar winds from AGB stars and SNe explosions are the main metal polluters, and~to track the metallicity evolution we introduce two parameters, following the example set by \cite{madau2014}: $R(z)$ represents the fraction of stellar mass returned to the ISM in the form of gas during stellar evolution, while $p(z)$ is the fraction of stellar mass that is turned into metals and ejected into the ISM. In~their work, ref. ~\cite{lacey2016} computed these parameters for a Kennicutt initial mass function (IMF). Here, we re-compute the values of these parameters for our fiducial IMF $\Phi(m)$, which corresponds to a Salpeter IMF within the $0.1-100 \mathrm{M_\odot}$ mass range, finding
\begin{equation}
R = \int_{0.1\;\!\mathrm{M_\odot}}^{100\;\!\mathrm{M_\odot}} \left(m - m_\mathrm{rem}(m) \right) \Phi(m)\; 
\mathrm{d}\ln m = 0.301
\end{equation}
and
\begin{equation}
p = \int_{0.1\;\!\mathrm{M_\odot}}^{100\;\!\mathrm{M_\odot}} p_Z(m)\;\! m\;\! \Phi(m)\; \mathrm{d}\ln m = 0.018,
\end{equation}
where $m_\mathrm{rem}(m)$ is the mass of the remnant (white dwarf, neutron star, or black hole) of a star of mass $m$ and $p_Z(m)$ is the fraction of the initial mass $m$ of a star that is synthesised into metal and ejected. Both $m_\mathrm{rem}(m)$ and $p_Z(m)$ are derived by interpolating the results from the stellar evolution calculations performed in~\cite{marigo1996,portinari1998}. The~metal production rate \mbox{then reads}
\begin{equation}
\dot{M}_Z = p \dot{M}_*.
\end{equation}
The new metals are injected into the cold gas mass component and~are then proportionally exchanged between the cold gas, hot gas, stellar, and~gas reservoir mass components.

\subsection{Black Hole Growth and Feedback~Model}
\label{bhacc} 
As presented in detail in PPW24, our black hole accretion and feedback model depends on the host halo mass. Below~the critical value $M_\mathrm{h}^\mathrm{crit}$, black holes can accrete only from the hot gas at the Bondi rate. Above~$M_\mathrm{h}^\mathrm{crit}$, major mergers (with mass ratio $\geq 0.1$) trigger episodes of fast accretion of cold~gas:
\begin{itemize}
    \item in the sEDD scenario, the accreted gas mass is limited to a fraction $\mathrm{f_{av}^{bh}}$ \mbox{(see Appendix \ref{appendixA})} of the total cold gas mass present in the galaxy;
    \item in the EDDlim scenario, the accreted gas mass corresponds to the minimum between the same fraction, $\mathrm{f_{av}^{bh}}$, of the cold gas mass or the mass that the black hole would accrete by growing at the Eddington rate.
\end{itemize}
For both scenarios, the~duration of the episode persists until the fraction of the cold gas mass, $m_\mathrm{c}$, is reduced to a proportion, $f_\mathrm{c}$, in relation to its value at the time of the merger. We write the mass accreted by the black hole at each time step as
\begin{equation}
\dot{M}_\mathrm{bh} = \begin{cases}
\dot{M}_\mathrm{bh}^\mathrm{hot} & 
  \text{for $M_\mathrm{h} < M_\mathrm{h}^\mathrm{crit}$}\\
\dot{M}_\mathrm{bh}^\mathrm{hot}+\dot{M}_\mathrm{bh}^\mathrm{cold} & 
  \text{for $M_\mathrm{h} > M_\mathrm{h}^\mathrm{crit}$}
\end{cases}\,. 
\end{equation}
The inflow of hot gas towards the central black hole, on~the other hand, is permitted continuously.

\textcolor{black}{As specified in PPW24, it is assumed that radiative feedback lunched during cold gas accretion results in gas outflow and replenishes the gas reservoir, while jets, if~they exist, can heat up the cold gas in the galaxy}. The~quasar luminosity, $\mathrm{L_{bol}}$, is calculated according to the numerical simulation of the slim discs of~\cite{sadowski2009} and fitted by~\cite{madau2014}. The~dimensionless black hole spin parameter is adopted to be $a = 0.5$ for all black holes. The~jet can be launched when either the condition of the accretion rate, in~terms of the Eddington ratio $\lambda$, is satisfied: $\lambda_\mathrm{Edd} \leq \lambda_\mathrm{ADAF,jet} = 0.01$ or $\lambda_\mathrm{Edd} \geq \lambda_\mathrm{slim,jet} = 1$. To~estimate the electromagnetic extraction of black hole rotational energy as Blandford--Znajek jet power, 
we consider the theoretical estimation in~\cite{tchekhovskoy2011}:
\begin{equation}
P_\mathrm{jet} = 2.8\;\! f(a)\left(\frac{\phi}{15}\right)^2 \dot{M}_\mathrm{bh}\;\! c^2\,.
\label{jet_power}
\end{equation}
The above formula is related to the dimensionless magnetic flux, $\phi$, and~the black hole spin dependence, $\mathrm{f}(a) = a^2\left(1+\sqrt{1-a^2}\right)^{-2}$. For~each black hole seed, a~random value is chosen between $\phi=1$ and $\phi=50$. In~the event of a black hole merger, the~resulting $\phi$ corresponds to that of the black hole possessing the greatest mass. After~the jet is launched, part of the jet energy, $f^\mathrm{h}_\mathrm{jet}$, heats up cold gas:
\begin{equation}
\dot{M}_\mathrm{heated} = f^\mathrm{h}_\mathrm{jet} \frac{2\;\! P_\mathrm{jet}}{{V_\mathrm{vir}}^2}.
\label{mgheated}
\end{equation}

%*******************************************************************
%%%%%%%%%%%%%%%%%%%%%%%%%%%%%%%%%%%%%%%%%%

\section{The Galaxy and SMBH~Populations}
\label{results_1}

To guarantee that the model thoroughly represents the evolution of both galaxy and AGN populations through cosmic time, in~this section we compare our results with observations. At~the same time, we provide an overview of the main characteristics of the jetted AGN populations as forecasted by our~model.

\subsection{Star Formation and Black Hole Accretion~Rates}
\begin{figure}[H]
%\centering
\includegraphics[width=0.75\textwidth]{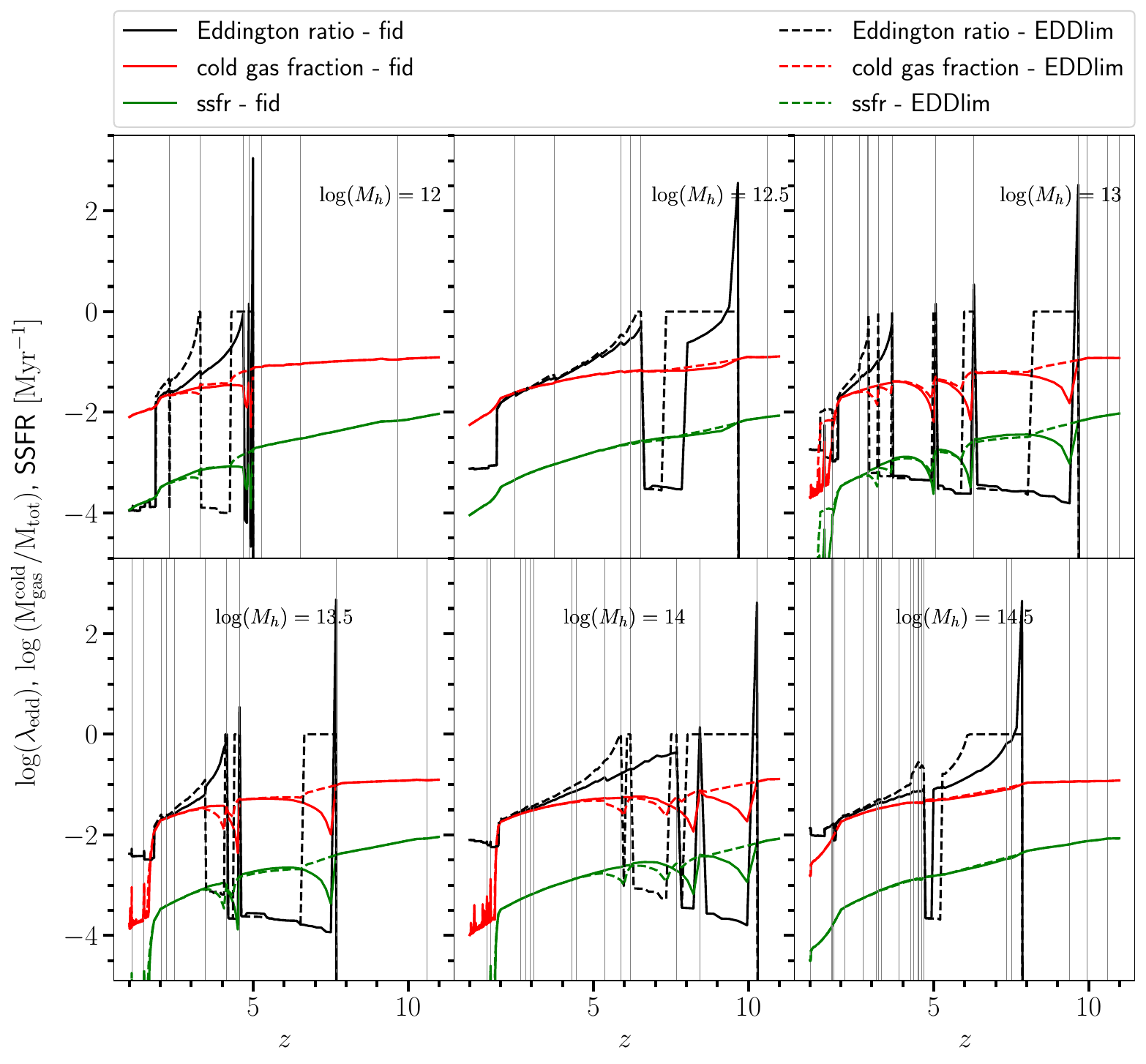}
\caption{Redshift %MDPI: We moved figure here to behind its citation, please confirm the revision.
 evolution for the cold gas fraction and specific star formation rate of galaxies, the~Eddington ratio of their central SMBHs for halos of different masses. Grey vertical lines represent major~mergers.}\label{jet_history}
\end{figure}

Figure~\ref{jet_history} shows the redshift evolution of the Eddington ratio, the~cold gas fraction, and the specific star formation rate (in $\mathrm{Myr^{-1}}$) for galaxies in halos of different masses. In~particular, SMBHs in the sEDD model show short, early strong slim-disc jet bursts with $\lambda_\mathrm{Edd} > 1$ that can temporarily quench both star formation and SMBH activity by heating the cold gas, until~major mergers trigger new episodes. Later on, as~the Eddington ratio decreases, the~SMBH enters a prolonged ADAF jet phase with $\lambda < \lambda_\mathrm{ADAF,jet} = 0.01$. On~the other hand, SMBHs in the EDDlim case initially grow slower and galactic gas accretion from the IGM is able to compensate the feedback from the black hole and sustain continuous star formation activity in the early evolutionary phases. In~this scenario, slim-disc jet bursts can last up to $\sim$100 Myr and~can be triggered multiple times by major mergers, before~$\lambda < \lambda_\mathrm{ADAF,jet}$ is satisfied and the SMBH enters the ADAF jet phase. In~this case, we can have jet-induced quenching episodes at later epochs. For~all considered quantities, the~differences between the two scenarios become negligible at later epochs, and~we expect this to reflect when looking at low-redshift population-wide statistical~observables.

\begin{figure}[H]
%\centering
\includegraphics[width=0.8\textwidth]{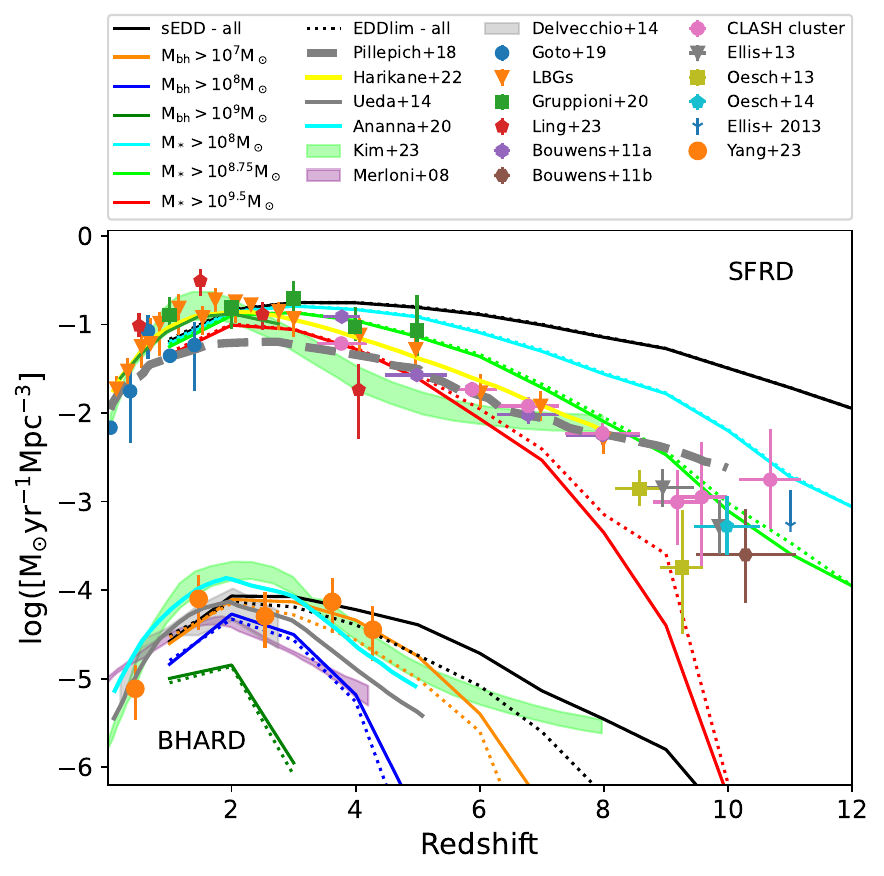}
\caption{Cosmic %MDPI: We moved figure here to behind its citation, please confirm the revision.
 SMBH accretion rate density and star formation rate density as a function of redshift for different BH \textcolor{black}{and} stellar mass cuts as specified in the legend. Solid lines represent the super-Eddington (sEDD) scenario, while dashed lines are for the Eddington-limited (EDDlim) case. Comparisons with several other observational and theoretical results are shown \protect\citep{pillepich2018, harikane2022, ueda2014, ananna2020, kim2023, merloni2008, delvecchio2014, goto2019, gruppioni2020, ling2023, bouwens2011a, bouwens2011b, ellis2013, oesch2013, oesch2014, yang2023}.}\label{sfrd}
\end{figure}

Figure~\ref{sfrd} shows the redshift evolution of the cosmic star formation and black hole accretion rate densities (SFRD, BHARD) for different mass cuts. Overall, our results match the observations across the whole redshift range considered, the~best agreement being achieved, respectively, for~galaxies with $M_* > 10^{8.75} \mathrm{M_\odot}$ and black holes with $M_\mathrm{bh}>10^7 \mathrm{M_\odot}$. We also reproduce the correct redshift for the peak of stellar and black hole activity, at~z$\sim$2. In~our model, this is caused by the combined effect of AGN jets emitted by SMBHs whose growth is now slowing down, with~the Eddington ratio falling below $\lambda_\mathrm{Edd} \leq 0.01$. These jets are capable of heating up most of the gas present in the galaxy. The~second effect is that high-mass galaxies accrete most of the gas from the IGM in hot~mode.

\subsection{Black Hole Mass~Function}
\begin{figure}[H]
\includegraphics[width=0.75\textwidth]{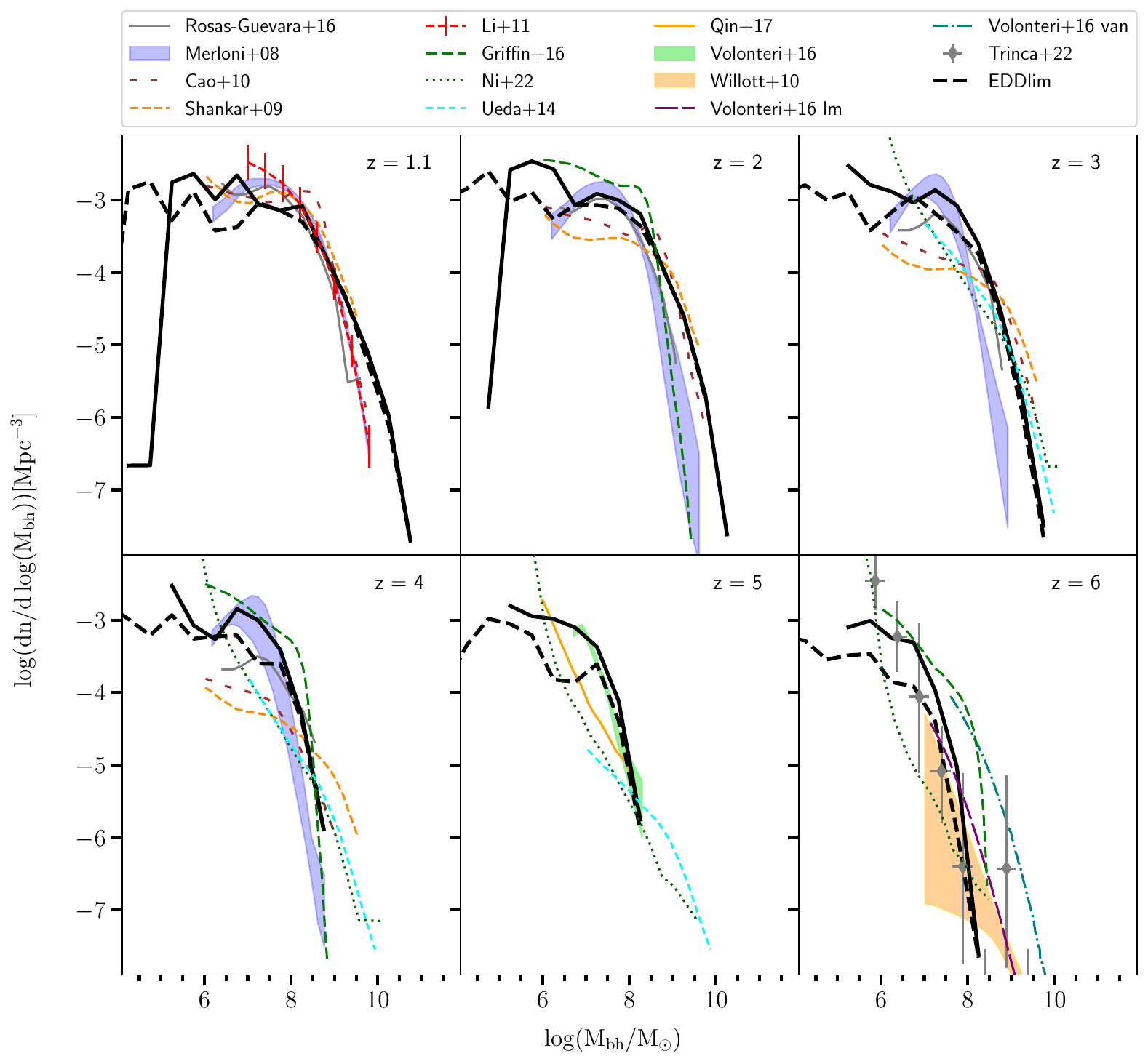}
\caption{Redshift %MDPI: We moved figure here to behind citation, please confirm the revision.
 evolution of the black hole mass function for super-Eddington (sEDD, black solid lines) and the Eddington-limited (EDDlim, black dashed lines) cases. We also show the results of other observational and theoretical results for comparison \protect\citep{rosasguevara2016, merloni2008, cao2010, shankar2009, li2011, griffin2018, ni2022, ueda2014, qin2017, volonteri2017, willott2010, trinca2022}.}\label{bhmf}
\end{figure}

Figure~\ref{bhmf} displays the redshift evolution of the black hole mass function (BHMF) for both the sEDD (solid lines) and EDDlim (dashed lines) scenarios. In~PPW2024, we showed that the two SMBH growth models produce markedly different SMBH populations in $z \gtrsim 7$, although~they exhibit a tendency to converge as the redshift decreases. In~this study, we validate this trend down to $z=1$ and illustrate that both models are in agreement with the observational and theoretical findings of the BHMF at $z <$ 6--7. 

\subsection{Jet Activities During Cosmic~Time}
\begin{figure}[H]
\includegraphics[width=0.75\textwidth]{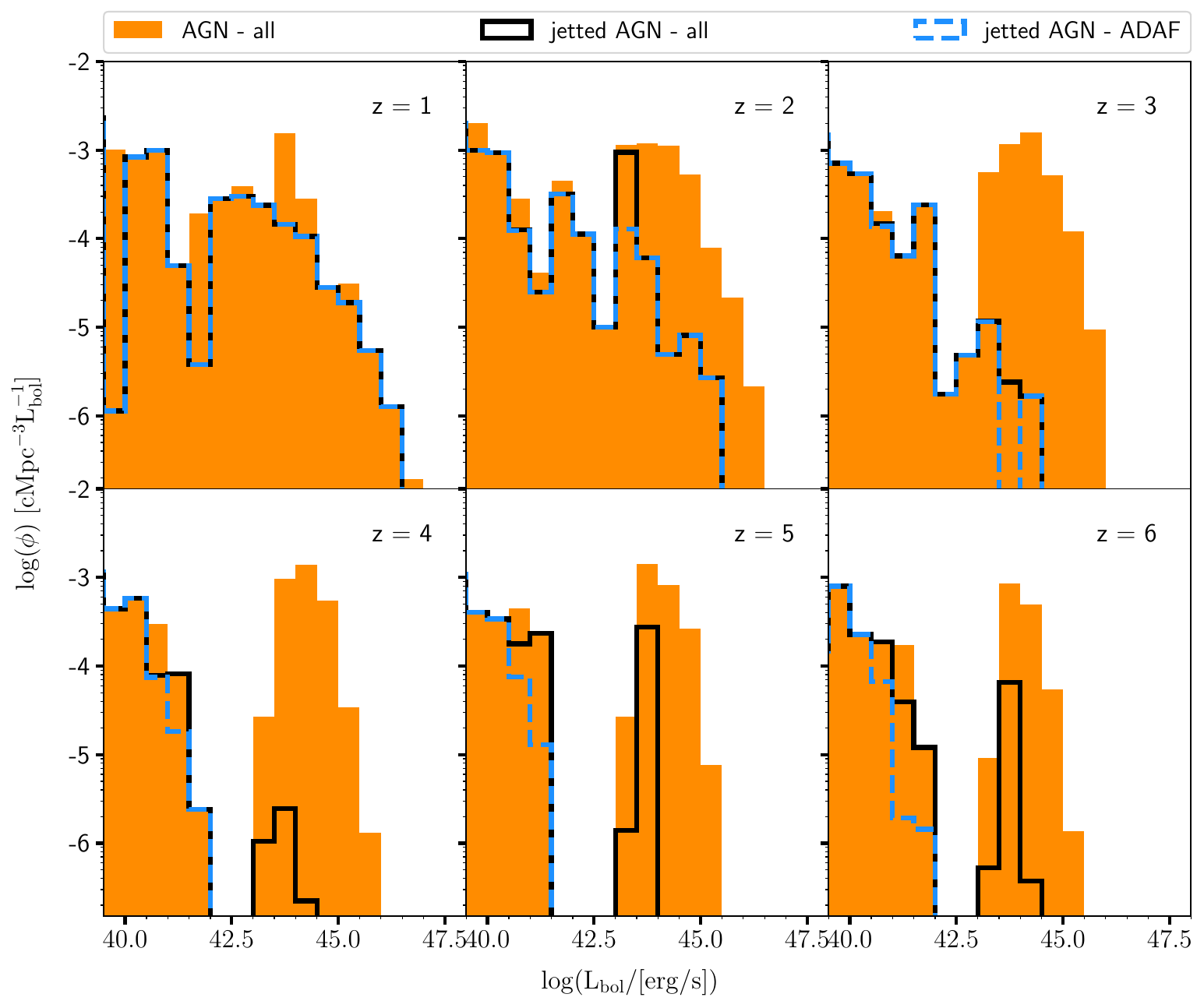}
\caption{The %MDPI: We moved figure here to behind its first citation, please confirm the revision. 2. Please change the hyphen (-) into a minus sign ($-$, "U+2212"). e.g., "-1" should be "$-$1".
 AGN bolometric luminosity function, $\phi$, across different redshifts for the sEDD model: it represents the expected number of AGNs found in a $\mathrm{Mpc^3}$ per each bin of $\mathrm{L_{bol}}$. The~histograms filled in orange represent the AGN population, while the black solid lines and blue dashed lines denote the jetted AGN and ADAF-jetted populations, respectively.}\label{AGN_blf}
\end{figure}

To examine the contribution of AGN jet activities, we compare AGN with jets (jetted AGN) to the entire AGN population. The~AGN bolometric luminosity function, $\phi$, based on the radiative output of the accretion disc, $L_\mathrm{qso}$, is shown for our benchmark scenario (sEDD model) in Figure~\ref{AGN_blf}. Among~jetted AGN, we further highlight that the population with their jet contribution belongs to ADAF jets. Note that the total jetted AGN populations consists of both sources with ADAF jets and slim-disc~jets.

In general, jet activities become increasingly prevalent in high-luminosity bins as redshift decreases, whereas, at high redshift, the~most luminous sources are associated with SMBHs in quasar mode, which characteristically lack jets. Slim-disc jets, originating from SMBHs undergoing super-Eddington accretion, predominate among high-luminosity jetted AGN at elevated redshifts. However, over~time, there is a decrease in the average Eddington ratio of SMBHs, a~trend corroborated by observational surveys. In~particular, at~z$\sim$1, no SMBHs are accreting at super-Eddington rates. Furthermore, the~gap observed in the luminosity function at high redshift is indicative of the bimodal distribution of the Eddington ratio, as~previously identified in Figure~6 of~PPW2024.

\begin{figure}[H]
%\centering
\includegraphics[width=0.75\textwidth]{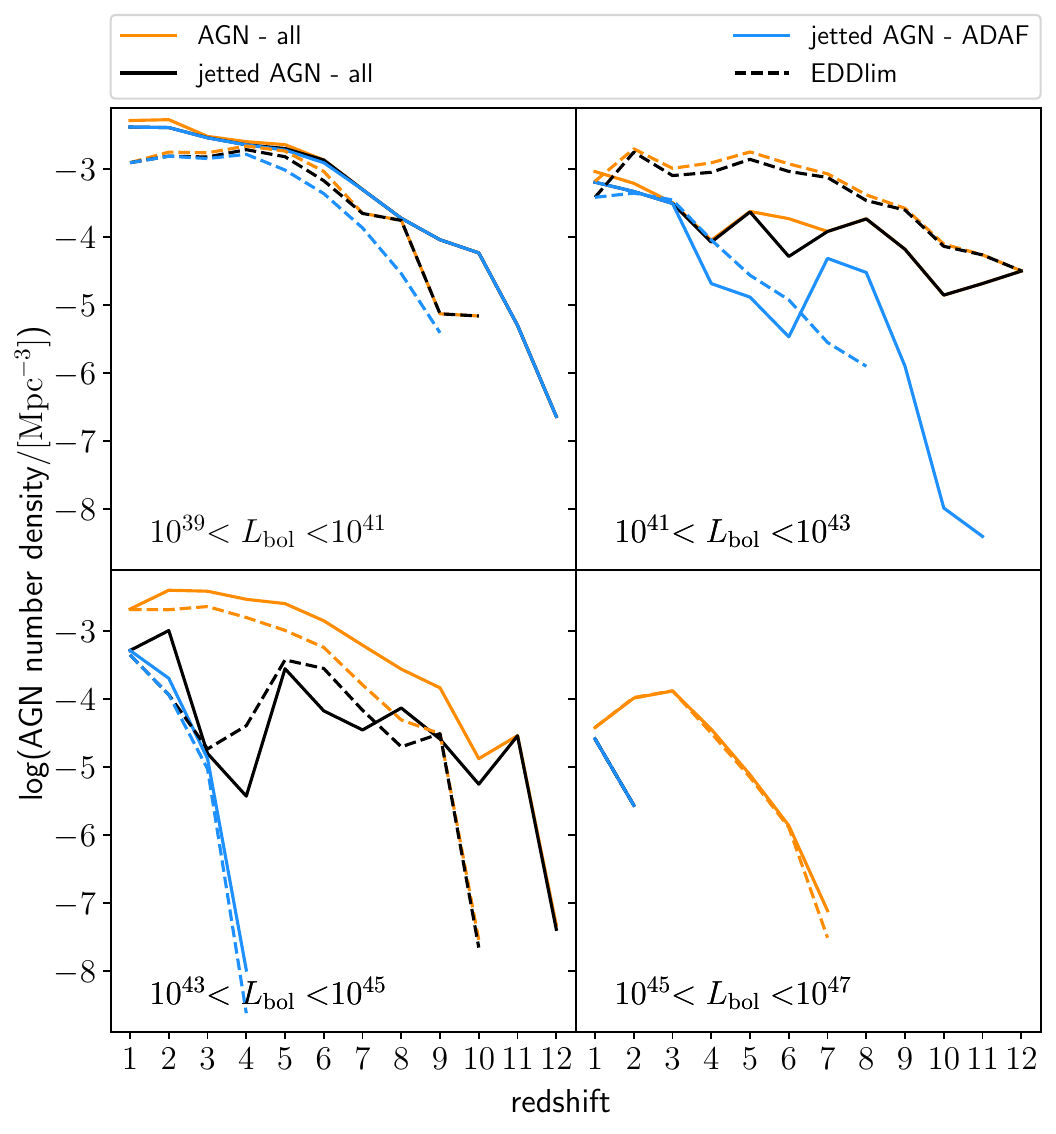}
\caption{Evolution of the number densities of the total (solid blue line, jetted (dashed orange line) and~ADAF$-$jetted (dotted green line) AGN populations as a function of redshift for different luminosity~bins.}\label{jetted_frac}
\end{figure}

The AGN number densities as a function of redshift for different populations (AGN, all jetted AGN, and~jetted AGN with ADAF jets) are presented in Figure~\ref{jetted_frac}. 
The evolution of the AGN number density is shown for different luminosity bins and for both the sEDD (solid lines) and the EDDlim (dashed lines) models, respectively. We note that the trends between the two growth models
differ significantly at $z \gsim 7$, due to the different growth paths SMBHs undergo in the early stages of their evolution, as~already shown in Figure~\ref{jet_history}. Jetted AGN dominate at low luminosities across all redshifts, while quasar-mode SMBHs are prevalent in the highest-luminosity bin and for $10^{43} < L_\mathrm{bol} < 10^{47}$ erg/s at $z \lsim$ 6--7.

\unskip   

\section{UHECR Source Spectra Associated with Jet~Activities}
\label{results_2}

To approach the question of how jet activities in different black hole growth models affect the evolution of UHECR background, we need to estimate the spectral energy distribution density of cosmic rays escaping from the jetted AGN systems and ejected into intergalactic space. We follow the work of~\cite{ptuskin2013} (see, e.g., \citep{eichmann2018} for a similar approach) \mbox{and define}
\begin{equation}
\label{eq:CR_rate}
    q(E) = \xi_\mathrm{CR} n_\mathrm{jet} P_\mathrm{jet} E^{-2} \mathrm{\Theta}\left(E_\mathrm{max} - E\right) \ ,
\end{equation}
where $\mathrm{\Theta}$ is the full Heaviside step function, the~parameter $\xi_\mathrm{CR}$ is the fraction of the total jet power that goes into accelerating particles, and $n_\mathrm{jet}$ is the jet number density. In~the formula, $E_\mathrm{max}$ is the maximum accelerating energy of a particle computed according to the Hillas criterion, and~reads~\cite{ptuskin2013}
\begin{equation}
\label{eq_cr}
    E_\mathrm{max} = \mathcal{Z}e\left(6 \beta c^{-1} P_\mathrm{jet}\right)^{1/2} \approx 2.7 \times 10^{20} \mathcal{Z} \beta \left(P_\mathrm{jet,45}\right)^{1/2} \mathrm{eV} \ ,
\end{equation}
where $P_\mathrm{jet,45} = P_\mathrm{jet}/(10^{45} \mathrm{erg\ s^{-1}})$, $\mathcal{Z}e$ is the particle charge, and $\beta$ is the jet velocity in units of the speed of light 
(see, e.g., \citep{lovelace1976, blandford1993, farrar2009}). 
For a more thorough discussion of the shape of the injected CR spectrum, see also \cite{ptuskin2003}, where the distribution function for the energetic particles behind SN shocks is~derived. 

To incorporate the potential different UHECR production rates in jets with ADAF or slim disk, we further consider $\xi_\mathrm{CR}$ in Equation~(\ref{eq:CR_rate}) for two different types of jet, as~$\xi_\mathrm{CR, ADAF}$ and $\xi_\mathrm{CR, slim}$. \textcolor{black}{We adopt $\beta = v_\mathrm{jet}/c \approx 1$ and $\mathcal{Z} = 1$ ($\mathcal{Z} = 8$), assuming a pure Hydrogen (Oxygen) composition.}
\textcolor{black}{The selected values presented in $\mathcal{Z}$ are intended to account for the uncertainties regarding the particle composition of UHECR (e.g., \citep{schroeder2019}).  Recent measurements appear to challenge pure Hydrogen compositions, instead favoring multi-species models, especially at high energies \citep{heinze2016, liu2016}.}

\begin{figure}[H]
\includegraphics[width=\textwidth]{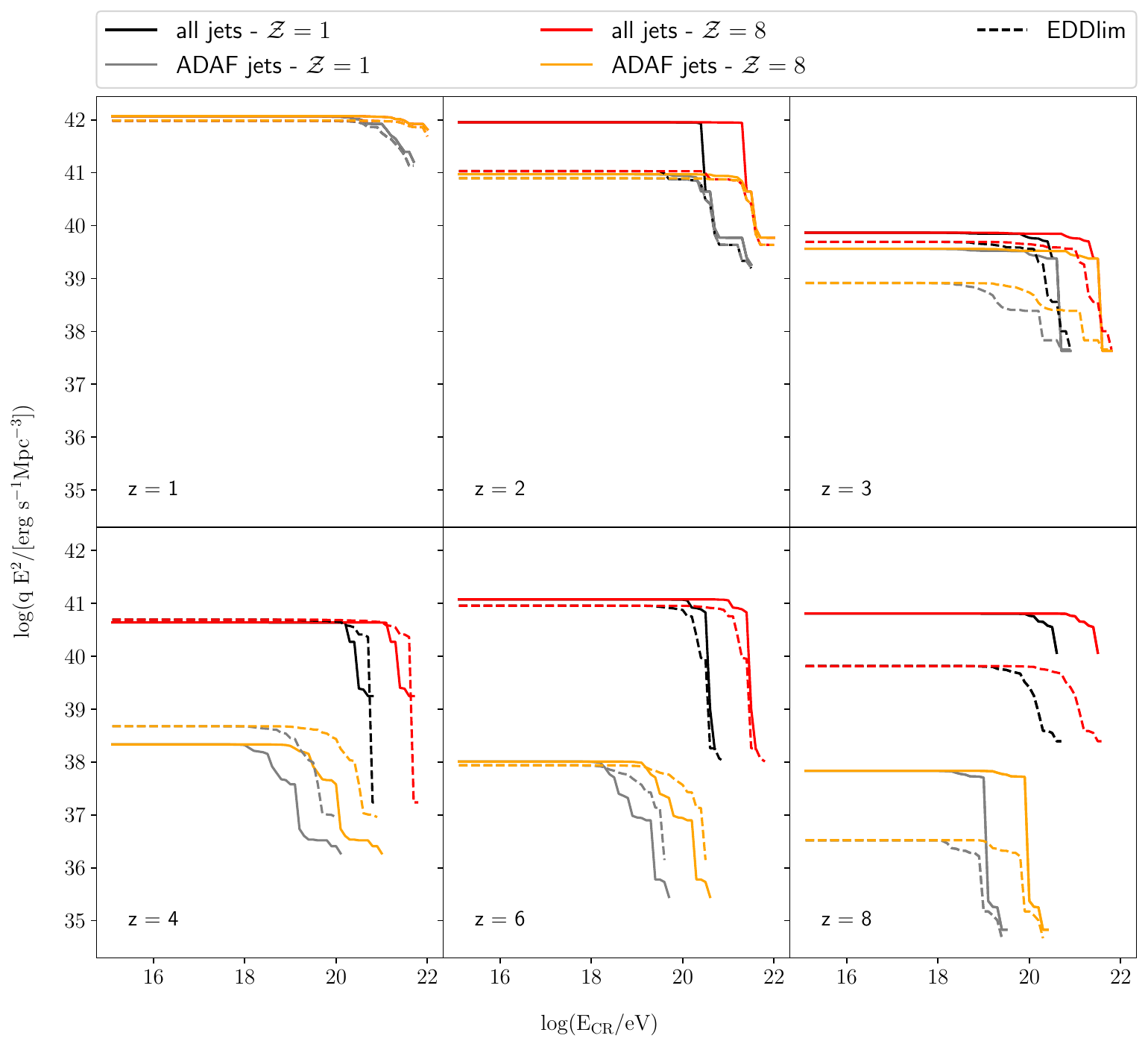}
\caption{Evolution of the high-energy cosmic ray spectral energy density emitted at different redshift in the sEDD (dashed lines) and EDDlim (solid lines) scenarios. \textcolor{black}{The results corresponding to different values of the atomic number $\mathcal{Z}$---assuming pure Hydrogen (black and gray lines) or Oxygen (red and orange lines) compositions---are presented.} The contribution to the total flux coming from ADAF jets, i.e.,\ for $\lambda_\mathrm{E} \leq 0.01$, as~opposed to the slim-disc jets, defined for $\lambda_\mathrm{E} \geq 1$, is also~shown.}\label{cr_spec}
\end{figure}

Figure~\ref{cr_spec} shows the results for the evolution of the total emitted CR energy density for the super-Eddington (sEDD, solid lines) and Eddington-limited (EDDlim, dashed lines) models at different redshifts. Two instances of UHECR production rates are considered: (1) UHECRs are produced by all jetted AGN, with~$(\xi_\mathrm{CR,~ ADAF},\xi_\mathrm{CR, slim})=(0.1, 0.1)$, as~indicated by the black (for $\mathcal{Z}=1$) and red (for $\mathcal{Z}=8$) lines in the figure, and~(2) UHECRs are produced only by the ADAF jet, with~$(\xi_\mathrm{CR,~ ADAF},\xi_\mathrm{CR, slim})=(0.1, 0)$, as~indicated by the grey (for $\mathcal{Z}=1$) and yellow (for $\mathcal{Z}=8$) lines in the figure. 
As expected, Oxygen (and generally heavier) nuclei would dominate the high-energy end of the spectrum. In~addition, this analysis demonstrates that, when considering only the influence of ADAF jets, the~energy density of the emitted UHECRs at $z \gsim 4$ corresponds to only 0.1--1$\%$ of the energy that would be emitted by slim-disc jets. This phenomenon arises because, as~previously observed, slim-disc jets originating from super-Eddington sources predominantly prevail at high $z$, while ADAF jets are more common at lower redshift. \textcolor{black}{The variance in UHECR spectra generated by ADAF jets between the sEDD and EDDlim models approximates an order of magnitude at z$\sim$3, whereas no discernible discrepancy is observed at $z<3$. Conversely, when considering contributions from all jets, a~difference of one order of magnitude is evident in the spectra emitted by the two models at z$\sim$2. The~distinctions between the sEDD and EDDlim models become more pronounced at $z\gtrsim 8$.} 

Due to the uncertainties of the adopted parameters, it should be noted that our primary focus is not in the magnitude of the energy fluxes illustrated in Figure~\ref{cr_spec}, but~rather in elucidating the relative differences between the various models. Under~the simplifying assumptions of this investigation, critical parameters such as jet power and the UHECR production rate, $\xi_\mathrm{CR}$, are postulated. In~fact, we assume that the jets are powered by magnetic extraction of the black hole rotational energy, and~therefore their power depends on the black hole spin and the magnetic flux surrounding it, both of which are parameters within our SMBH growth models PPW2024.
More precise predictions could be achieved in the future with enhanced understanding of the physics related to these free parameters. Most of the models that track the evolution of the SMBH spin (e.g., \citep{fanidakis2011, dubois2014, bustamante2019}) follow the formalism developed in~\cite{bardeen1970}. However, the~resulting spin distribution strongly depends on the assumed BH accretion model and on the initial conditions. A~more conservative approach, in~this sense, would be considering upper and lower limits to the cumulative jet power by setting all BH spin parameters equal to, e.g.,~0.15 and 0.998 \citep{fanidakis2011}. According to Equation~(\ref{jet_power}), this would result in a factor of $\approx$150 difference between the upper and lower limits of the jet~power.

\textcolor{black}{It is crucial to recognize that CRs experience energy loss during their propagation. Among~the factors contributing to attenuation between their site of production and Earth, photohadronic ($p\gamma$) processes, particularly photo-pair production, are predominant in the energy loss of UHECRs. The~interaction between UHECRs and cosmic microwave background photons establishes the Greisen--Zatsepin--Kuzmin (GZK) horizon \citep{greisen1966, zatsepin1966}, which restricts UHECRs from reaching Earth from distant sources as they dissipate energy prior to arrival. Moreover, the~propagation length of cosmic rays decreases with increasing redshift (see, e.g., \citep{dermer2010,wu2024}), suggesting that the disparity in the UHECR spectrum sEDD and EDDlim models at high redshift could instead manifest itself in variations within other particle constituents as they degrade in energy. For~example, our framework allows for the estimation of the diffuse neutrino background contributed by AGNs, for~given models of neutrino production within AGN jets (e.g., \citep{jacobsen2015}).}

%%%%%%%%%%%%%%%%%%%%%%%%%%%%%%%%%%%%%%%%%%
\section{Summary and~Implications}
\label{conclusions}
In this study, we explore and demonstrate the role of AGN jets as potential sources of extragalactic UHECRs. We present the cosmological semi-analytic framework, \textit{JET}, to~model the evolution of jetted AGN populations and their connection to UHECR fluxes. Our framework provides a comprehensive approach to understanding the formation and evolution of galaxies, SMBHs, and~their associated jet phenomena from \( z = 20 \) to \( z = 1 \). By~incorporating jet formation mechanisms, black hole accretion states, and~feedback processes into a cosmological context, \textit{JET} enables predictions of the diffuse UHECR densities under different SMBH growth~models.

Our results highlight several key insights for the connection between UHECR and AGN jets, as~itemized~below.
\begin{itemize}
\item \textcolor{black}{Dependence on accretion types}:
the nature of AGN jets and their efficiencies are strongly influenced by the accretion states of SMBHs, as~determined by their Eddington ratios. The~UHECR production rate for the different jets (and therefore the accretion types), as~considered in our parameter $\xi_{\rm CR}$, is a major source of uncertainty.
\item \textcolor{black}{Dependence of jet duty cycles and feedback}:
the duty cycles of AGN jets, shaped by SMBH accretion histories, significantly impact the predicted UHECR flux at the source. Feedback effects further regulate the evolution of SMBHs and their host galaxies, leading to differences in the emitted cosmic ray energy densities. 
\item \textcolor{black}{Dependence of SMBH growth models}: different SMBH growth scenarios, such as the
sEDD and EDDlim models considered in this work, lead to varying predictions for 
the UHECR~background. 

\end{itemize}

Furthermore, since the production of neutrinos and gamma rays is linked to the UHECR processes, such as proton--proton ($pp$) or photohadronic ($p\gamma$) interactions, the~implications of our findings are not solely restricted to the theoretical predictions concerning cosmic rays but also encompass neutrino and gamma ray fluxes. In~particular, the~relative flux between UHECR, neutrinos, and~gamma rays is associated with the prevailing interactions. \textcolor{black}{Although the high redshifts of UHECR are unable to reach Earth owing to attenuation, the~assessment of the diffuse neutrino background originating from AGN jets can be furthered through an understanding of neutrino production within these jets.} This study reveals a direct connection between cosmological models of SMBH growth, AGN jet formation, and~astroparticle backgrounds, thereby providing a framework for interpreting observational data in the realm of multi-messenger astrophysics. Further theoretical inquiries are necessary to explore these dependencies more~comprehensively.

%%%%%%%%%%%%%%%%%%%%%%%%%%%%%%%%%%%%%%%%%%
%% optional
%\supplementary{The following supporting information can be downloaded at:  \linksupplementary{s1}, Figure S1: title; Table S1: title; Video S1: title.}


%\begin{tabularx}{\textwidth}{lll}
%\toprule
%\textbf{Name} & \textbf{Type} & \textbf{Description} \\
%\midrule
%S1 & Python script (.py) & Script of python source code used in XX \\
%S2 & Text (.txt) & Script of modelling code used to make Figure X \\
%S3 & Text (.txt) & Raw data from experiment X \\
%S4 & Video (.mp4) & Video demonstrating the hardware in use \\
%\bottomrule
%\end{tabularx}
%}

%%%%%%%%%%%%%%%%%%%%%%%%%%%%%%%%%%%%%%%%%%
\vspace{6pt}

\authorcontributions{Conceptualization, O.P. and H.-Y.P.; methodology, O.P. and H.-Y.P.; software, O.P.; validation, O.P.; formal analysis, O.P.; investigation, O.P. and H.-Y.P.; writing---original draft, O.P.; writing---review and editing, O.P. and H.-Y.P.; visualization, O.P.; supervision, H.-Y.P.; funding acquisition, H.-Y.P. All authors have read and agreed to the published version of the manuscript.}

\funding{This work is supported by the Yushan Fellow Program of the Ministry of Education (MoE) of Taiwan (ROC), 
the National Science and Technology Council (NSTC) of Taiwan (ROC) under the
grant 112-2112-M-003-010-MY3. %MDPI: Please add: ``This research received no external funding'' or ``This research was funded by NAME OF FUNDER grant number XXX.'' and  and ``The APC was funded by XXX''. Check carefully that the details given are accurate and use the standard spelling of funding agency names at \url{https://search.crossref.org/funding}, any errors may affect your future funding.
}
% Only for journal Nursing Reports
%\publicinvolvement{Please describe how the public (patients, consumers, carers) were involved in the research. Consider reporting against the GRIPP2 (Guidance for Reporting Involvement of Patients and the Public) checklist. If the public were not involved in any aspect of the research add: ``No public involvement in any aspect of this research''.}

% Only for journal Nursing Reports
%\guidelinesstandards{Please add a statement indicating which reporting guideline was used when drafting the report. For example, ``This manuscript was drafted against the XXX (the full name of reporting guidelines and citation) for XXX (type of research) research''. A complete list of reporting guidelines can be accessed via the equator network: \url{https://www.equator-network.org/}.}

% Only for journal Nursing Reports
%\useofartificialintelligence{Please describe in detail any and all uses of artificial intelligence (AI) or AI-assisted tools used in the preparation of the manuscript. This may include, but is not limited to, language translation, language editing and grammar, or generating text. Alternatively, please state that “AI or AI-assisted tools were not used in drafting any aspect of this manuscript”.}

\dataavailability{The data underlying this paper, which have been produced by our model, are available on reasonable request to the corresponding author.} 

\acknowledgments{The %MDPI: please confirm if the funding information in the Acknowledgments Section should be moved to the Funding Section.
 authors express their gratitude to Pratika Dayal%MDPI: Titles (e.g., Dr., Mr., and Prof.) should NOT be used in the Acknowledgments section. We removed them. Please confirm.
, who initially developed the DELPHI semi-analytic model, from~which our model is derived, and~to the entire DELPHI group in Groningen. Additionally, the~authors are grateful to Kinwah Wu for the enlightening discussions on cosmic ray physics that significantly contributed to the definition of the study's focus. The~authors also acknowledge the participants of the workshop ``Interstellar and Intergalactic Insights: Exploring the Energetic Universe with Multi-Messengers'' (held at National Taiwan Normal University, 2023) for their valuable discussions on Astroparticle Physics. The~authors also wish to thank the careful referees who helped in improving the paper. This work has made use of the NASA Astrophysics Data System.}

\conflictsofinterest{The authors declare they have no conflicts of interest with respect to this~study.} 

\appendixtitles{no} % Leave argument "no" if all appendix headings stay EMPTY (then no dot is printed after "Appendix A"). If the appendix sections contain a heading then change the argument to "yes".
\appendixstart
\appendix
\section[Appendix A]{}\label{appendixA}
For completeness of information, we list in Table~\ref{table_params} the free parameters of the model, tuned to reproduce the statistical AGNs and galaxy observables shown in the paper. See PPW2024 for more details on their defining equations. $\Delta_z = [\mathrm{\Omega_m}(1+z)^3+\mathrm{\Omega_\lambda}]^{1/3}$.

\begin{table}[H]
\caption{Model parameters, description, and default~values.\label{table_params}}
\begin{tabularx}{\textwidth}{cCc}
\toprule
\textbf{Parameter} & \textbf{Description} & \textbf{Value}\\
\midrule
$f_\mathrm{cold}$ & fraction of the gas mass accreted onto the galaxy as cold & 0.4\\
$f_\mathrm{*}$ & star formation efficiency cap & 0.02\\
$f_*^\mathrm{w}$ & fraction of SN energy that couples to the gas & 0.1\\
$M_\mathrm{h}^\mathrm{crit}$ & critical halo mass for BH growth and cold accretion & $10^{11.25} {\Delta_z}^{-3/8} \mathrm{M_\odot}$\\
$\mathrm{M_{mm}}$ & halo mass ratio defining major mergers & 0.1\\
$f_\mathrm{av}^\mathrm{bh}$ & fraction of cold gas mass that BH can accrete & 0.00003\\
$f_\mathrm{c}$ & limiting cold gas fraction for quasar accretion & 0.6\\
$f_\mathrm{qso}^\mathrm{w}$ & fraction of BH energy that couples to the gas & 0.003\\
$f_\mathrm{jet}^\mathrm{w}$ & fraction of jet energy that drives outflows & 0.003\\
$f_\mathrm{jet}^\mathrm{h}$ & fraction of jet energy that heats up the gas & 0.01\\
\noalign{\smallskip}
\bottomrule
\end{tabularx}
\end{table}

%%%%%%%%%%%%%%%%%%%%%%%%%%%%%%%%%%%%%%%%%%
%% Optional

%% Only for journal Encyclopedia
%\entrylink{The Link to this entry published on the encyclopedia platform.}

%%%%%%%%%%%%%%%%%%%%%%%%%%%%%%%%%%%%%%%%%%
\begin{adjustwidth}{-\extralength}{0cm}

\reftitle{References}



%=====================================
% References, variant B: internal bibliography
%=====================================

%%%%%%%%%%%%%%%%%%%%%%%%%%%%%%%%%%%%%%%%%%
%% for journal Sci
%\reviewreports{\\
%Reviewer 1 comments and authors’ response\\
%Reviewer 2 comments and authors’ response\\
%Reviewer 3 comments and authors’ response
%}
%%%%%%%%%%%%%%%%%%%%%%%%%%%%%%%%%%%%%%%%%%
\PublishersNote{}
\end{adjustwidth}
\end{document}